# Finite size and inner structure controlled by electrostatic screening in globular complexes of proteins and polyelectrolyte†

**Jérémie Gummel,**[a] **François Boué,**[a] **Daniel Clemens,**[b] **and Fabrice Cousin**[a]*

5 *a Laboratoire Léon Brillouin, CEA Saclay 91191 Gif sur Yvette Cedex France*
*b Hahn Meitner Institut, BENSC, Glienicker Straße 100, 14109 Berlin-Wannsee Germany*

We present an extended structural study of globular complexes made by mixing a positively charge protein (lysozyme) and a negatively charged polyelectrolyte (PSS). We study the influence of all the parameters that may act on the structure of the complexes (charge densities and concentration of the species, partial hydrophobicity of the polyion chain, ionic strength). The structures on a
15 scale range lying from 10Å to 1000Å are measured by SANS. Whatever the conditions, the same structure is found, based on the formation of dense globules of ~ 100Å with a neutral core and a volume fraction of organic species (compacity) of ~ 0.3. At higher scale, the globules are arranged in fractal aggregates. Zetametry measurements show that globular complexes have **a** total positive charge when the charge ratio of species introduced in the mixture [-]/[+]intro > 1 and a total
20 negative charge when [-]/[+]intro < 1. This comes from the presence of charged species in slight excess in a layer at the surface of the globules. The globule finite size is determined by the Debye length $\kappa^{-1}$ whatever the way the physicochemical parameters are modified in the system, as long as chain-protein interactions are of simple electrostatics nature. The mean number of proteins per primary complex $N_{lyso\_comp}$ grows exponentially on a master curve with $\kappa^{-1}$. This enables to picture
25 the mechanisms of formation of the complexes. There is an initial stage of formation where the growth of the complexes is only driven by attractions between opposite species associated with counterion release. During the growth of the complexes, the globules progressively repell themselves by electrostatic repulsion because their charge increases. When this repulsion becomes dominent in the system, globules stop growing and behave like charged colloids: they aggregate
30 with a RLCA process, which leads to the formation of fractal aggregates of dimension Df 2.1.

## Introduction

The field of complexation of polyelectrolyte and proteins of opposite charge has developed a lot these last years, as shown by the numerous reviews published on the subject dealing
35 with polyelectrolytes – protein systems[1,2] or more specifically on polysaccharides – proteins[3] or simulations.[4] This is first due to the numerous applications in which these systems are involved such as drug release,[5] biochips[6] or protein fractionation.[7] It is also driven by the new polyelectrolyte
40 architectures such as polyelectrolyte multilayers[8] or polyelectrolyte dendrimers[9] which now allow protein immobilization on flat surfaces[10] or on colloidal particles.[11-13]

† Electronic Supplementary Information (ESI) available: [
45 Measurement of the sulfonation rate of PSS chains, fitting of SANS experiments made at pH 3, experimental SANS data and fitting of experiments made at pH 7, pH 4.7 and f = 0.5, pH 4.7 whith a variation of the ionic strength of the buffer and pH 4.7 when the lysozyme concentration is varied.].
50 See DOI: 10.1039/b000000x/

This addresses the question of the mechanisms of the complexation: the main interaction in the system clearly
55 seems to be the electrostatic one, as now well established in literature, apart from simultaneous less important effects, which we recall now. First, it appears that hydrophobic interactions do not play a dominant role except in specific cases. An example is when the polyelectrolyte backbone bears
60 pending alkyl chains.[14,15] Another one is a reorganization stage that occurs after an initial electrostatic interaction,[16] and eventually leads to protein unfolding.[17] Second, the intrinsic stiffness of the chains also naturally acts on the complexation: the binding of spherical macroions on polyelectrolytes
65 decreases with an increase of the chain stiffness, as shown by turbidity measurements[18] on micelles/polyelectrolytes and proteins/polyelectrolyte systems, in agreement with simulations.[19,20] But these effects remain generally of the second order compared to electrostatic ones.
70 Focusing thus on electrostatics from now on, the key parameters in the system that will mostly influence the structure of the complexes are thus the charge density of the protein, the charge density of the polyelectrolyte and the ionic strength of the system that screens the electrostatic interactions on the typical range of the Debye length $1/\kappa$. They rule several processes:

-The critical conditions required for complexation at room temperature are closely linked with these 3 parameters because surface charge density controls the adsorption/desorption limits that is proportional to $\kappa^{-1}$.[21] The screening of the electrostatic interactions via $\kappa^{-1}$ is a key point since simulations[20,22] and experiments[23] have shown that when the ionic strength becomes too high, the attraction between polyelectrolytes and proteins is screened enough to prevent the formation of complexes. Compared to colloids or micelles, proteins add a degree of complexity for the description of the electrostatic interactions since their surface charge can be heterogeneous owing to the presence of patches of different charges at the surface of protein.[23] Both simulations[22] and experiments[24,25] have shown that the presence of heterogeneous charges enhances the complexation.

- Once complexation occurs, the aggregation is maximum when the stoichiometry of charges brought in solution by the two species is equal to 1,[24-28] in agreement with simulations.[29] Focusing on a more detailed example, such stoichiometry is fulfilled inside the core of a complex, while proteins in electrostatic excess can be free or polyelectrolyte chain in excess are localized in a corona.[28]

- There is a competition between enthalpic and entropic effects, leading to either exothermic[16,30] or endothermic[27] complexation as measured using calorimetry, but the key parameter seems to be the charge density of the polyelectrolyte. Highly charged polyelectrolytes can indeed gain a lot of entropy during the complexation due to release of condensed counterions as proposed[31,32] and experimentally checked.[33]

Electrostatics is clearly present also when it comes to the structure of the complexes, but several spatial effects and different scales have also to be considered:

- on local scales (10 to 1000 Å) literature reports two main kinds of structures, with quite different density: the system can form either gels [14,17,34,35], or dense globular aggregates of a few hundreds of Å. [24,26,27,17,28] At this stage, the spatial scale of the chain size is relevant. The globular structure is common to the many similar systems dealing with polyelectrolyte and charged spheres of opposite charge such as micelles[35,36] or inorganic spheres.[37] It is also obtained by simulations[29,38] that show in particular that the compacity of the clusters increases with the surface charge density of the spheres.[38] Experiments on both proteins/polyelectrolyte[39] or micelles/polyelectrolyte systems[40] show it is determined by the dilution state of the polyelectrolyte chains: gels are formed when chains can make bridges whereas dense globular complexes are formed otherwise. In reference[39] the threshold between the two regimes corresponds exactly to the overlapping concentration of the chains c* if one accounts for prior interaction with the proteins, which modifies the global size of the chains.

- On a higher scale (> 1000 Å), on the reverse, the system evolves, usually slowly, in particular when it is made of dense globules; this evolution depends on the electrostatic charge density of the objects. Highly charged systems tend to precipitate to form a fractal structure of globules[41,42] whereas coacervation occurs for less charged systems.[35,43] But concerning scales and electrostatics in this globular regime, despite all these recent advances, several questions, which are crucial for the understanding of the mechanisms of the complexation, are still not answered. Above all, the central question is: what is tuning the finite size –and the compacity- of the primary complexes? Do charge patches and hydrophobicity play a role, or is it all just electrostatics, and in this case what is the exact process?

In this paper, we answer these questions on a given model system by testing the influence of all the parameters that may act on the structure (charge densities and concentration of the species, partial hydrophobicity of the polyion chain, ionic strength). This model system is a mixture of lysozyme, a globular positive charged protein and Polystyrene Sulfonate (PSS), a flexible highly negative charged polyelectrolyte (PSS), which we chose because: (i) it enables to tune all the parameters depicted before independently and (ii) it provides powerful possibilities of contrast matching for Small Angle Neutron Scattering (SANS) experiment by using deuterated PSS chains. We will indeed mainly use SANS among all the structural techniques available because this technique has enabled us to describe with an unequalled precision the structures of the lysozyme/PSS complexes from local scales to larger scales for one specific set of physicochemical parameters (pH fixed, hydrophobicity of the chain fixed and ionic strength of the buffer fixed).[17,28,39,41] For such parameters, and when the ratio between the positive charges brought by the protein and the negative charges brought by the polyelectrolyte ($[-]/[+]_{intro}$) is close to 1, we can encounter the two main structures of complexes evoked above: above c* overlapping concentration of the chains, we encounter a gel-like aqueous material inside which water swells a network of PSS chains crosslinked by the protein. Below c*, we find dense globular complexes of about 100 Å, which we call "primary complexes", arranged in a fractal way with a 2.1 dimension up to micron scales below c*. We have previously shown, in particular, that the core of the primary globular complexes is neutral and the species in excess from an electrostatic point of view remain free in solution.

We extend here our SANS experiments in the globular complex regime, which we complete with zetametry experiments, on a wide range of conditions to measure their radius, effective charge and compacity. We use 3 pH values (3, 4.7 and 7) in order to vary the net charge of the lysozyme. This will in particular enable us to compare the situations where lysozyme essentially bears positive charges (pH 3 and pH 4.7) to the situation where it bears both positive and negative charges (pH 7).[44] We use PSS chains with two rates of sulfonation f of 1 and 0.5. This enables to change both the linear charge density of the PSS chain and its hydrophobicity because partially sulfonated chains present a partial hydrophobic behavior.[45] We tune the ionic strength of the buffer from 50 mM to 500 mM. We work with small PSS chains (50 repetitions units) keeping the chain length constant. This enables to work with a high content of components during the SANS experiments to maximize the scattering because the PSS chains stay in diluted regime after interaction with lysozyme up to 0.4 mol/L for an ionic strength of 0.5 mol/L. We show that this displays a unified behavior of this regime, observed for the first time to our knowledge, for this kind of protein-polyelectrolyte systems.

## Experimental

### Sample preparation

The sulfonation of polystyrene chains is done by ourselves following a method derived from the Makowski method.[46,47] It is done on deuterated polystyrene chains of 50 repetitions units with a very low polydispersity (Mw/Mn ~ 1.03) purchased from Polymer Standard Service. Depending on the ratio of the quantity of reactive species to the quantity of polystyrene chains during the sulfonation reaction, the value of the ratio of monomers that are sulfonated during the reaction can be tuned in a range between 40% and 100% to get the poly(styrene-co-styrene sulfonic acid). This post-sulfonation is statistic. We have done two lots of chains with respective sulfonation rates of $f = 0.5$ and $f = 1$. The sulfonation rate has been accurately checked by SANS (see supporting information). In the following, both partially and totally sulfonated chains will be called PSS chains. Lysozyme is purchased from Sigma and used without further purification.

In order to tune the pH of the mixture three different buffers solutions have been used: chloroacetic acid/chloroacetate ($CH_2ClCOOH/CH_2ClCOO^-$) at pH 3, acetic acid/acetate ($CH_3COOH/CH_3COO^-$) at pH 4.7 and monobasic phosphate/dibasic phosphate at pH 7 ($H_2PO_4^-/HPO_4^{2-}$). The counterions are $Na^+$ sodium in all cases. The buffer concentration is set by default to $5.10^{-2}$ mol/L. For specific experiment where we use buffer with higher ionic strength, NaCl is added to one of the three buffers to reach a range of ionic strength lying between $5.10^{-2}$ mol/L and $5.10^{-1}$ mol/L. We also prepared a specific sample at $5.10^{-1}$ mol/L with an acetate only buffer.

All samples are done in the same way. Two solutions, one of lysozyme and one of PSSNa, are first prepared separately in a given buffer and then mixed and slightly shaken to be homogenized. The samples are then left at rest for two days. We checked in previous experiments that this is enough to reach a state which shows no significant change over days or even months, such that we have enough time to achieve all our measurements. Apart from specific experiments described in the corresponding sections, we use always a lysozyme concentration of 40g/L to get a good SANS signal. The PSSNa concentration is then adapted to make samples with charge ratios denoted $[-]/[+]_{intro}$ lying from 0.5 to 3.33. $[-]/[+]_{intro}$ is the rate of negative charges to positive charges of species introduced initially in the mixture. It takes into account structural charge and not effective charge. It is thus calculated with the net charge of lysozyme (+ 17 at pH 3, +11 at pH 4.7 and +8 at pH 7) and with one negative charge per sulfonated PS monomer on the PSS chains. As soon as mixing is done, a very turbid fluid solution is obtained in all cases.

### SANS experiments

SANS measurements were done either on V4 spectrometer (HMI, Berlin, Germany) or on the PACE spectrometer (LLB, Saclay, France) in a q-range lying from $3.10^{-3}$ to $3.10^{-1}$ Å$^{-1}$. All measurements were done under atmospheric pressure and at room temperature.

In order to get either the PSSNa signal or the lysozyme signal independently, each PSS/protein compositions was achieved in two solvents: once in a fully $D_2O$ buffer that matches the neutron scattering length density of deuterated PSSNa, and once in a 57%/43% $H_2O/D_2O$ mixture that matches the neutron scattering length density of lysozyme . Standard corrections for sample volume, neutron beam transmission, empty cell signal subtraction, detector efficiency, subtraction of incoherent scattering and solvent buffer were applied to get the scattered intensities in absolute scale for complexes samples.

### Zetametry

Zetametry measurements were done on a home-made apparatus that uses a technique of microelectrophoresis in liquid vein. The sample is introduced in a thin tube with electrodes at both ends to apply the electric field. A laser illuminates the sample and the light is backscattered by the complexes, which enables their tracking. The speed of the complexes is measured with a microscope. The distribution of the speed of the particles is obtained by averaging the path of every complex tracked on the microscope. In our measurements, this distribution was narrow enough to get a good average speed and thus of an average charge of the particles.

Because our samples are very turbid and very concentrated, a prior dilution of the samples (by a factor 1000) is necessary to allow tracking of individual objects. Other experiments show us that the possible reorganizations of the system only occur on scales higher than the primary complexes, i.e only within the fractal network of primary complexes.[48] The size measured after dilution indeed lies between 500 Å and 1000 Å, which means that the aggregation number of complexes $N_{agg}$ has decreased toward a few dozen of primary complexes whereas it was initially higher than several hundreds.[41] Since all samples in this paper have a similar structure (see section 3), we believe that the reorganizations occur in a same way for all samples. Therefore the sizes here cannot be compared with to the values of $N_{agg}$ obtained from SANS before dilution. The zetametry would thus provide us only relative measurements of the ζ potential from one sample to another one. The absolute measurement of the ζ potential is not possible because $N_{agg}$ after dilution is not precisely measured.

## Results

We will first describe the structure of the primary complexes obtained, which is the same for all the conditions tested in the paper (change of pH, sulfonation rate, ionic strength, concentration of species). This structure is thus central; it has been described in a previous work of ours for a

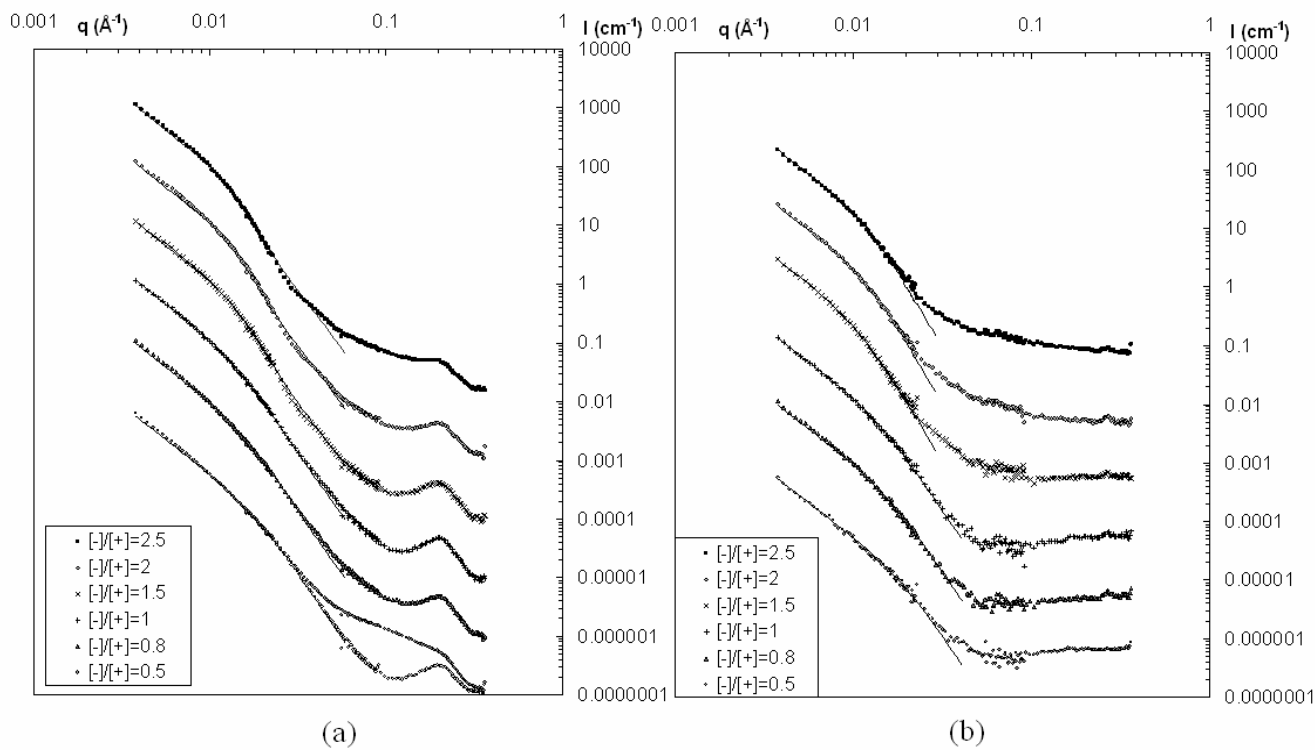

**Fig. 1** samples at pH 3, f = 1, I = 5 10$^{-2}$ and [Lyso]$_{intro}$ = 40g/L: (a) lysozyme scattering. All curves are shifted from one to another by a decade for clarity. The scattering for the [-]/[+]$_{intro}$ = 0.5 is presented with or without the subtraction of free lysozyme. The intensity in absolute scale correspond to the [-]/[+]intro = 2.5 sample. The full lines correspond to the fits. (b) PSS chains scattering. All curves are shifted from one to another by a decade for clarity. The intensity in absolute scale correspond to the [-]/[+]$_{intro}$ = 2.5 sample. The full lines correspond to the fits. The errors bars are smaller than the symbols.

first limited set of experimental conditions (pH 4.7, f = 1, I = 5 10$^{-2}$ and [Lyso]$_{intro}$ = 40g/L).[28,41] We recall that it corresponds to globular complexes with a core for which we can measure the inner composition, hence the inner charge ratio [-]/[+]$_{intro}$. We will then consider the compacity and most of all the global size of the complexes.

Finally we give their outer charge, which involves zetametry measurements. These last two parameters will be the core of our discussion.

**Structure of the complexes from SANS measurements when varying electrostatic interactions**

**A globular structure**

Whatever the experimental conditions, we recover the main features of the scattering spectra described in former papers,[28, 41] both when matching proteins signal or PSS signal. We show here only the scattering results obtained at pH 3 as an example (Fig. 1). Other curves are in the Supplementay Information, for sake of clarity. The spatial organization, obtained from a modeling method we have extensively described in reference[28], is made at large scale of fractal aggregates (2.1 dimension) composed at lower scale of globules of ~10 nm. They are made of a dense core, which are surrounded, when [-]/[+]$_{intro}$ > 1, by a thin shell of PSS chains. When [-]/[+]$_{intro}$ ≤ 1, there are free proteins in solution. When [-]/[+]$_{intro}$ >> 1, there are free PSS chains in solution.

We present in SI all the quantitative information on the primary complexes obtained from the modelization of the curves: the mean radius $R_{lyso}$ obtained when only lysozyme scattering is measured, the mean radius $R_{PSS}$ obtained when only PSS scattering is measured, and the **inner** volume fraction of lysozyme (resp. PSS) inside the globular complexes $\Phi_{lyso\_inner}$ (resp. $\Phi_{PSS\_inner}$).

**Inner charge ratio**

Before presenting the overview on all experiments, let us first show how the fine use of contrast variation in SANS is a powerful method to get, on primary complexes, a direct information on the inner charge ratio independently from the fitting procedure. This is simple when the primary complexes are 'naked', i.e without any PSS shell at the surface, for [-]/[+]$_{intro}$ ~1. In this case the size of the primary complex is very similar from the lysozyme point of view and from the PSS point of view. The scattered intensity writes then:[28]

$$I_{comp\ lys}(q)\ (cm^{-1}) = \Phi_{comp}\Delta\rho_{comp\ lys}^2 V_{comp}P_{comp}(q)S_{comp}(q) \quad (1a)$$

$$I_{comp\ PSS}(q)\ (cm^{-1}) = \Phi_{comp}\Delta\rho_{comp\ PSS}^2 V_{comp}P_{comp}(q)S_{comp}(q) \quad (1b)$$

where $\Phi_{comp}$ is the volume fraction of primary complexes, $V_{comp}$ is their volume, $P_{comp}$ their form factor and $S_{comp}$ their structure factor. The effective contrast between complexes and solvent where lysozyme is visible (PSS matched), $\Delta\rho_{comp\ lys}^2$ equals $\Delta\rho_{lys}^2 \cdot \Phi_{lys\_inner}^2$ (we repeat that $\Phi_{PSS\_inner}$ is the **inner** volume fraction of lysozyme inside the globular complexes. The same is true replacing lysozyme by PSS in the symetric matching. Due to the symetry, $\Delta\rho_{PSS}^2 = \Delta\rho_{lys}^2$, and as $\Phi_{comp}$, $V_{comp}$, $P_{comp}(q)$ and $S_{comp}(q)$ are similar for the two species at low q, one gets:

$$I_{PSS}(q)/I_{lyso}(q) = (\Phi_{PSS\_inner}/\Phi_{lyso\_inner})^2 \quad (2)$$

One can then define $[-]/[+]_{inner}$, the rate of negative charges to positive charges of species present inside the globules, which is proportional to $\Phi_{PSS\_inner}/\Phi_{lyso\_inner}$.

From Eq. (2), $I_{PSS}(q)/I_{lyso}(q)$, the ratio of the scattering of PSS chains to the scattering of lysozyme, is thus proportional to the square of the inner charge ratio, $[-]/[+]_{inner}{}^2$. It is thus a very accurate method for the measurement of the inner charge ratio. The scattering ratios are shown for the different cases for samples in figure 2.

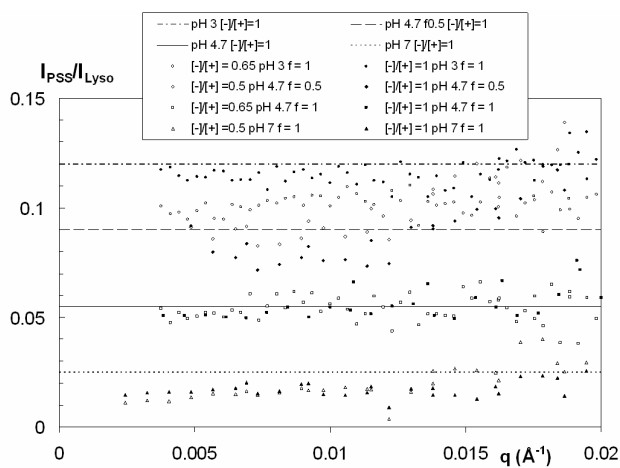

**Fig. 2** $I_{PSS}(q)/I_{lyso}(q)$ in the low q regime for all samples when $[-]/[+]_{intro} \leq 1$ (I = 5 $10^{-2}$ mol/L). Results for pH 4.7 and f = 1 are issued from a former paper.[28] The full lines correspond to $[-]/[+]_{inner}$ = 1. The errors bars are smaller than the symbols

Their experimental values are constant on a q-range lying from 0.003 Å$^{-1}$ to 0.02 Å$^{-1}$ and are very close to the continuous lines that correspond to the theoretical ratio if the inner charge ratio is $[-]/[+]_{intro}$ exactly equal to 1. Since the mean volume of a repetition unit is lower for f = 0.5 than for f = 1 (some repetitions units do not bear the sulfonate group), the factor between the two theoretical ratios at pH 4.7 is not 2.

For the 'hairy' complexes, $[-]/[+]_{inner}$ can no longer simply obtained by $I_{PSS}(q)/I_{lyso}(q)$. The values are in this case obtained from the fitting of SANS data because $R_{lyso}$ and $R_{PSS}$ are different. Since $R_{lyso}$ is the radius of the core, it is calculated by dividing the amount of negative charge within the core obtained from $\Phi_{PSS\_inner}$ and $R_{lyso}$ by the amount of positive charge obtained from $\Phi_{lyso\_inner}$ and $R_{lyso}$.

We consider in this section a unique ionic strength I = 5 $10^{-2}$ mol/L, at which we make an overview of the inner charge $[-]/[+]_{inner}$ within the complexes as a function of $[-]/[+]_{intro}$, in various cases of pH and sulfonation rate. The values of $[-]/[+]_{inner}$ obtained from the fitting of SANS data (Supporting Information) or from $I_{PSS}(q)/I_{lyso}(q)$ for the three new sets of charge density of components (pH 3 and f= 1, pH 7 and f= 1, pH 4.7 and f= 0.5) are presented in Fig 3 and are compared with the values obtained at pH 4.7 and f = 1 in the previous study.[28,41] **The very striking result of Figure 3 is that $[-]/[+]_{inner}$ is always close to 1, whatever $[-]/[+]_{intro}$, whatever the pH and whatever the partial hydrophobicity of the chains.**

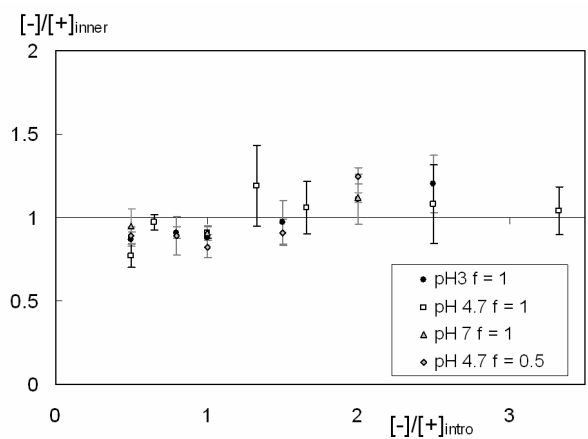

**Fig. 3** Overview of the variation of $[-]/[+]_{inner}$ (as extracted from comparison of Small Angle Neutron Scattering by lysozyme only and by PSS only) versus $[-]/[+]_{intro}$. Results for pH 4.7 and f = 1 are issued from a former paper.[28]

The stoichiometry of charges within the core of the primary complexes is recovered independently from the initial conditions. The sum of interactions is thus dominated by direct electrostatic attractions in the complexation process. Moreover, all data superimpose on a master curve when one takes into account $[-]/[+]_{intro}$ although the concentrations of PSS chains introduced during synthesis are very different from one pH to another (the concentration of lysozyme is constant). In all cases most of the species in excess from an electrostatic point of view remains free in solution. However there remains a slight excess of negative charges at low $[-]/[+]_{intro}$ or positive charges at large $[-]/[+]_{intro}$ in the inner of the complexes. This is a common behavior for all set of data, and the transition between the two regimes lies in a range of $[-]/[+]_{intro}$ varying from 1 to 1.5. We recall here that we calculate the charge ratio in the inner of the core of the primary complexes. For the 'naked' complexes without a PSS shell it is a direct determination of their charge but not for the 'hairy' complexes since the shell of PSS chains has to be taken into account.

It is remarkable that the main structures of the primary complexes are affected neither by the presence of an important number of negative charges on a globally positively charged protein which rises to 10 at pH 7 (compared to only 2 at pH 3) nor by the hydrophobic units of partially sulfonated chains. For pH 3 and pH 7, the main structures are finally similar to the ones presented previously[28] for pH 4.7. They are recalled in Fig. 9. For partial sulfonation, they slightly differ from the total sulfonation ones (see part 2.3 of S.I.): this could be due to the pearl necklace conformation, which has been observed in pure solution[49] and is still present in the complexes because pearls may induce a higher roughness at the surface of the complexes and thus perturb the formation of the shell of PSS. Finally, we note that electrostatics stoichiometry observed in Fig. 3 is consistent with the release of counterions from the core during complexation early predicted,[31,32] and which we reported recently from a direct scattering measurement.[33] In summary, at this stage, Fig. 3 is the first important result of this paper and supports the generality of electrostatics effects in these systems.

## Influence on the structure of the salinity of the buffer

We test in this section the influence of a higher salinity of the buffer on the structures of the complexes. We have prepared different samples at I = 1 $10^{-1}$ mol/L and I = 5 $10^{-1}$ mol/L at pH 4.7 with a given [-]/[+]$_{intro}$ (3.33). We compare their scattering in the Fig. 4 of the Supporting Information with the one obtained at I = 5 $10^{-2}$ mol/L in a previous study.[28] For I = 5 $10^{-1}$ mol/L, we have tested two buffers (a full solution of NaCH$_3$COO at I = 5 $10^{-1}$ mol/L and a mixture of NaCH$_3$COO with NaCl). Both samples give the same results, proving **that the nature of the counterion has a minor influence on the structure of the primary complexes**. The increase of the salinity of the buffer has nevertheless a huge influence on the complexes because it strongly increases their size. For the highest ionic strength, the mean complex radius $R_{mean}$ is too large to allow a satisfying fit and is at least 300 Å. But the electrostatic stoichiometry in the core of the complex is conserved. Please note that this increase of the size of the primary complexes has already been observed in a previous paper of ours.[39]

This formation of large complexes totally disagrees with the behavior obtained from simulations on polyelectrolyte/protein complexation[19,20] which predict that the strong screening of electrostatic interactions at large ionic strength no longer allow the complexation between charged colloids and polyelectrolytes. We can keep in mind that simulations generally deal with only one protein and one polyion; in the actual core, PSS chains and lysozyme are many. They are in close contact and form a lot of bonds. Moreover the exact nature of these bonds may depart from simple electrostatic bonds.

## Size and compacity of complexes when varying electrostatic interactions

Although direct electrostatic attractions between proteins and polyelectrolyte may lead in principle to aggregates of infinite size, we clearly observe complexes of well-defined finite size. In order to determine its origin, we make in this section an overview of the influence of all parameters that can influence this size (pH, [-]/[+]$_{intro}$, concentration of the species). The influence of the ionic strength of the buffer has already been presented in previous section (strong increase of the size of the complexes with an increase of ionic strength the buffer). All other results come from fits to SANS measurements displayed in Supporting Information. We consider here the mean diameter $R_{mean}$ of the core of the complexes obtained from the lysozyme scattering through the average volume:

$$R_{mean}^3 = R_{comp}^3 e^{\frac{9}{2}\sigma^2} \quad (3)$$

assuming a log-normal distribution of mean value $R_{comp}$ and variance $\sigma$. This average volume, knowing $\Phi_{lyso\_inner}$ from inner composition, enables to get the mean number of proteins per primary complex $N_{lyso\_comp}$:

$$N_{lyso\_comp} = \frac{4\pi}{3} \frac{R_{mean}^3 \Phi_{lyso\_inner}}{V_{lyso}} \quad (4)$$

Fig. 4.a summarizes the effect of [-]/[+]$_{intro}$ on $R_{mean}$ for all the different charge densities of components of the study : fully sulfonated chains for pH 3, 4.7 and 7 and partially sulfonated chains at pH 4.7. The lysozyme concentration is kept constant at 40g/L and the ionic strength at I = 5 $10^{-2}$ mol/L. Several trends can be noted:

- We recover the main conclusion of Refs[28,41] for all conditions: $R_{mean}$ strongly increases when [-]/[+]$_{intro}$ is increased. $N_{lyso\_comp}$ is typically 5 to 10 times higher at [-]/[+]$_{intro}$ = 2.5 than at [-]/[+]$_{intro}$ = 0.5.
- Complexes are slightly larger at pH 4.7 than at pH 3. They are much larger at pH 7. But in this last case the polydispersity is large (see table SI.2 in S.I.), which shifts $R_{mean}$ towards large values.
- When varying f (at pH 4.7) from 1 to 0.5, $R_{mean}$ is reduced (by a factor 1.5-2) but the reduction is enhanced for $N_{lyso\_comp}$ because complexes are less dense for f = 0.5.

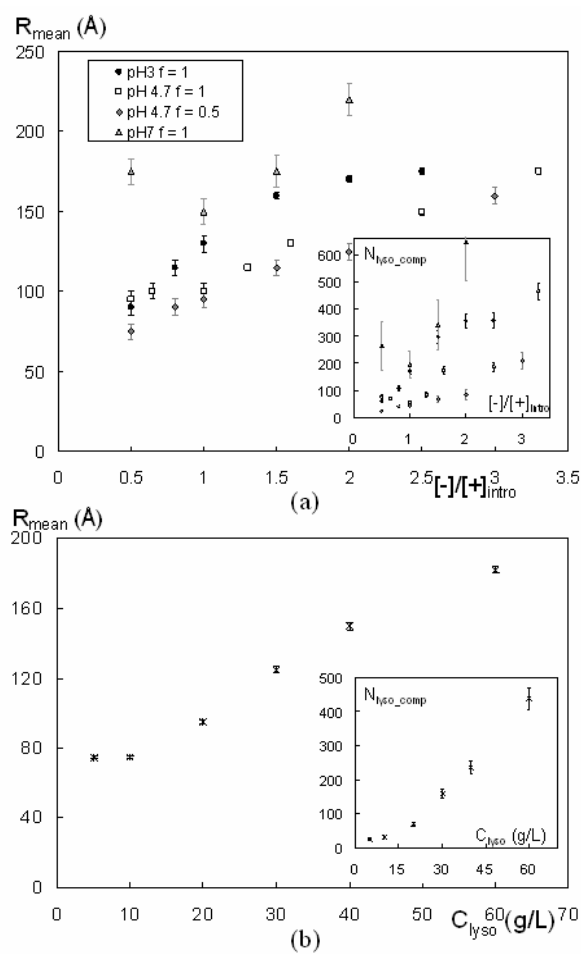

**Fig. 4** Size and number of proteins for the primary complexes: (a) $R_{mean}$ versus [-]/[+]$_{intro}$ ([lyso]$_{intro}$ = 40g/L, I = 5 $10^{-2}$). Insert: $N_{lyso\_comp}$ versus [-]/[+]$_{intro}$. (b) $R_{mean}$ versus concentration of species ([-]/[+]$_{intro}$ = 1.66, I = 5 $10^{-2}$). Insert: $Nl_{yso\_comp}$ versus concentration of species. The full line is a guide for the eye. Results for pH 4.7 and f = 1 are issued from a former paper.[28]

Fig 4.b summarizes the effect of the concentrations of species introduced on $R_{mean}$ for a constant $[-]/[+]_{intro}$ (1.66), keeping the ionic strength of the buffer constant (see results in S. I.):

- $R_{mean}$ shows an initial plateau, before to increase strongly with species concentration.
- the variation of $N_{lyso\_comp}$ (insert of Fig 4.b) follows the same trend; at low concentration of introduced species we see that $N_{lyso\_comp}$ is only a few tens. This behavior suggests an initial nucleation around germs. These small initial nuclei contain all the species in minority from the point of view of electrostatic ratio (protein if $[-]/[+]_{intro} < 1$, polyelectrolyte if $[-]/[+]_{intro} > 1$).

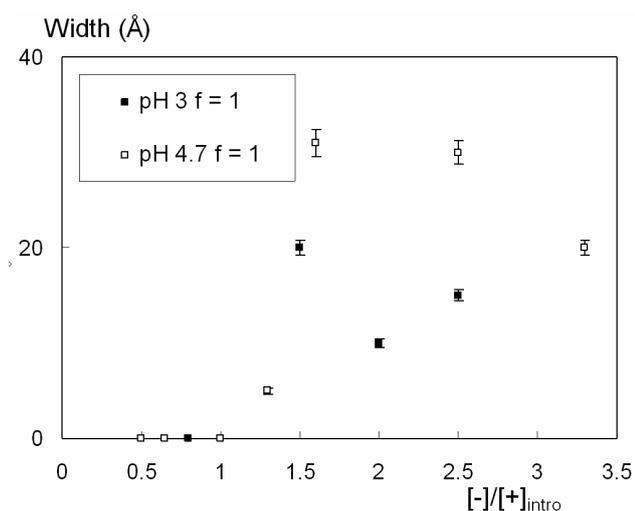

**Fig. 5** Width of the PSS shell surrounding complexes for fully sulfonated chains. Results for pH 4.7 and f = 1 are issued from a former paper.[28]

In summary, since all the physicochemical parameters have a direct influence on the finite size of the complexes, there is no main parameter arising at this stage, which would clearly explain the mechanisms for the final finite size.

The data fitting of the PSS scattering enables to get $R_{PSS}$ for totally sulfonated chains for both pH 3 (see table SI.1) and pH 4.7 (see table 1 of[41]). For pH 7 and for partially sulfonated chains at pH 4.7, the PSS scattering does not allow a correct fit of the data (see SI). The subtracting of $R_{lyso}$ from $R_{PSS}$ gives the width of the shell of PSS chains around complexes. It is presented in Fig. 5.

In accordance with the electrostatic stoichiometry of the core of the complexes, the shell is present only when $[-]/[+]_{intro} > 1$. It has a constant thickness at given pH: 20 Å at pH 3, 30 Å at pH 4.7, contrary to $R_{mean}$. This thickness can be related with the molar mass of the PSS chains: with 50 repetitions units of effective length 2.05 Å,[51] the fully extended chain should be 100 Å long. Whereas the persistence length may be short if the chain concentration in the shell is high, thickness values like 30 Å or even 20 Å suggest other effects. A large part of the chain could be deeply anchored within the core of the complexes. The anchoring should be higher at pH 3 than at pH 4.7, because the lysozyme bears a higher net charge. Another explanation is the existence of loops, which can exist if the persistence length is short enough. Note from our measurements of chain conformation[39] that we obtain for $L_p$ values as short as $L_p = 20$ Å for chains in the gel phase prior to globules, and that chains appear strongly shrunk in the globule phase.

The shell must be even thinner for partially sulfonated chains because $I_{PSS}(q)/I_{lyso}(q)$ very slightly increases towards low q for $[-]/[+]_{intro} > 1$ up to $[-]/[+]_{intro} > 2$ (see S.I.). It seems sensible that these chains should be less extended when they are less charged and partially collapsed in pearl necklaces due to hydrophobicity as proposed theoretically[52] and checked experimentally for pure solutions of partially sulfonated chains[49,50] that must be here buried under the surface. Due to the short mass of the chain, it is very difficult to get a pearl localized only in the shell.

The data fitting enables also to get the compacity of the primary complexes from $\Phi_{inner} = \Phi_{lyso\_inner} + \Phi_{PSS\_inner}$ (the rest of the matter inside the globules is water). Contrary to $\Phi_{lyso\_inner}$, $\Phi_{PSS\_inner}$ is not always fitted properly because the counting statistics are not good enough at low PSS concentration (see SI). When $\Phi_{PSS\_inner}$ cannot be extracted, we have estimated it by assuming that a neutral core ($[-]/[+]_{inner} = 1$) is formed in all conditions. In the latter case, the final error on $\Phi_{inner}$ stays very weak because $\Phi_{PSS\_inner}$ is always much lower than $\Phi_{lyso\_inner}$ ($\Phi_{PSS\_inner}/\Phi_{lyso\_inner}$ lies between ~ 0.15 at pH 7 for f = 1 and ~ 0.25 at pH 4.7 for f = 0.5). Figure 6 gives an overview of the compacity as a function of all relevant parameters (pH, f, $[-]/[+]_{intro}$, species concentration). Results at higher salinity of the buffer are missing: the fitting becomes impossible since the globules are too large. The main trend is that over all conditions, $\Phi_{inner}$ stays close to 30%(+/- 10%). This can be discussed as follows:

First of all, this proves that the increase of the size of the complexes with the increase of the concentration of species in solution corresponds mostly to an increase of the number of proteins (and of chains) within the primary complexes and not to a swelling of the primary complexes by the solvent. The independence of $\Phi_{inner}$ with $[-]/[+]_{intro}$ (Fig 6.a), at given pH, is consistent with the electrostatic stoichiometry: the inner structure, the arrangement of the same objects (chains and proteins), with same charge and same ratio, can be the same.

- At different pH, the objects are not the same: we see that $\Phi_{inner}$ decreases with an increase of pH: (around 35% at pH 3, 30% at pH 4.7 and 25% at pH 7): lysozyme is less charged, it possesses also more negative patches, in accordance with the simulations of Ulrich et al.[38]

- $\Phi_{inner}$ also slightly decreases with a decrease of the sulfonation rate ($\Phi_{inner} \sim 0.22$ for f = 0.5). In this case again the density of charge of one sign is lower (the PSS chains), so that they interact less with opposite ones. But this may also comes from pearls of partially hydrophobic PSS chains[49,50] which are buried within the core that (i) lower the accessibility of negative charges to the protein and (ii) does not allow a close packing. The compacity of the complexes is mainly due to proteins ($\Phi_{lyso\_inner}$ is much larger than $\Phi_{PSS\_inner}$) and can be considered as of the same order of magnitude than the compacity that would have been obtained in some proteins crystals (structures of proteins crystals obtained with different salts: Ries-Kautt et al.[53]).

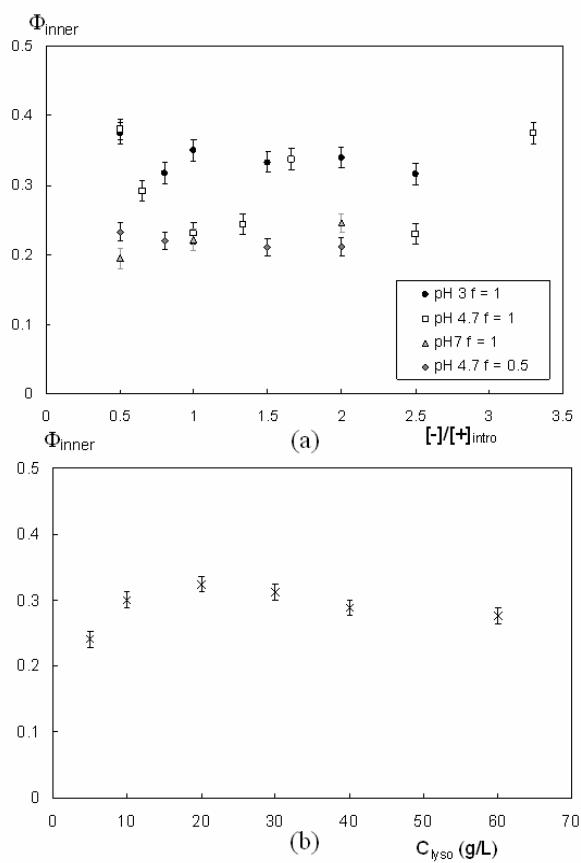

**Fig. 6** Overview of $\Phi_{inner}$ in the primary complexes. (a) Influence of the pH, the sulfonation rate and $[-]/[+]_{intro}$. (b) Influence of the concentration of the species introduced in the samples. Results for pH 4.7 and f = 1 are issued from a former paper.[28]

But in the globule, the order is only local, between first neighbours: the peak at 0.2 Å$^{-1}$ is never followed by other diffraction peaks at higher q in the lyzozyme scattering. The wrapping of PSS chains around proteins prevents crystallization. The high value of the compacity is again consistent with the electrostatic stoichiometry of the complexation. The PSS chains must indeed cover all the positive charges of lysozyme to reach such compacity.

**Effective charge of primary complexes: zetametry measurements.**

As noticed above and clear from Fig 3, $[-]/[+]_{inner}$ is slightly lower than 1 when $[-]/[+]_{intro} \leq 1$ and slightly higher than 1 when $[-]/[+]_{intro} > 1$. The primary complexes should thus bear a global charge, which would also explain why the solution remains macroscopically stable. Electrostatic repulsion prevents flocculation. In order to check this, we have performed zetametry measurements on complexes to determine their charge. We recall that the $\zeta$ potential of a given colloidal object measured in zetametry is an effective charge, lower than its structural charge, as it takes into account opposite charge brought by the counterions condensed at the surface of the object in the Stern layer.

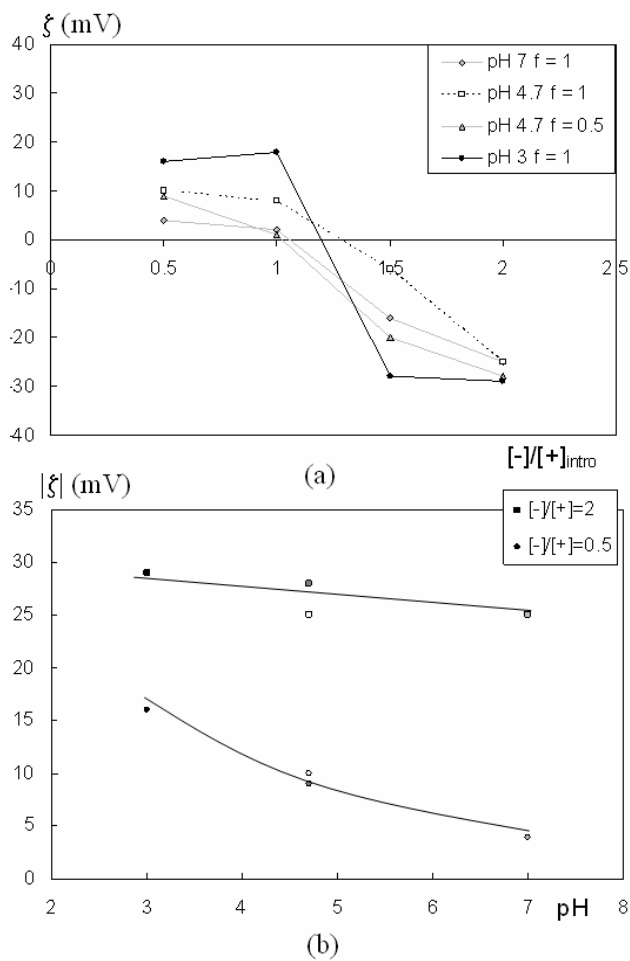

**Fig.7** $\zeta$ potential of primary complexes. (a) Measurement as a function of $[-]/[+]_{intro}$. (b) Absolute value $|\zeta|$ as a function of pH for $[-]/[+]_{intro} = 0.5$ and $[-]/[+]_{intro} = 2$.

Results presented in Fig. 7 agree quite well with the description driven from SANS, for samples with 4 different values of $[-]/[+]_{intro}$ in four different experimental conditions (pH 3, 4.7 and 7 for f =1 and pH 4.7 for f = 0.5). The results confirm that the primary complexes are positively charged when $[-]/[+]_{intro} \leq 1$ and negatively charged when $[-]/[+]_{intro} > 1$.

Beyond confirmation of SANS global charge analysis, **zetametry brings additional information, which are more obvious on Figure 7.b** which plots the absolute value $|\zeta|$ of the potential as a function of pH in the case of positive ($[-]/[+]_{intro} = 0.5$) and negative charge excess ($[-]/[+]_{intro} = 2$):

- First, $|\zeta|$ is always larger for negative charge excess (upper curve). This agrees with our expectation that the dangling PSS chains of the shell, charged on each segment, bring more charge excess than do lysozyme on the other side of the $[-]/[+]_{intro} = 1$ boundary. The difference, which depends on pH, is nevertheless not so strong between the two opposite sides $[-]/[+]_{intro} = 0.5$ and 2... It can be explained by the fact that a large fraction of counterions stay condensed on the chains within the shell, as visualized with labeled c.i. in a former study of ours.[33]

- Second, the variation with pH of $|\zeta|$ is also informative..

The negative charge for [-]/[+]$_{intro}$ > 2 hardly varies with pH and seems to tend to a value independent of pH. This is akin to the behavior of a PSSNa unit in a usual solution, where it is a strong acid. We obtain the same value of |ζ| at pH 4.7 for fully and partially sulfonated chains. This can mean that the dissociation of PSSNa units is complete in both case, but the total charge is regulated by counterions condensation.This therefore agrees with SANS which tells us there is a shell of polymer, though this is not an additionnal proof. On the opposite side, for [-]/[+]$_{intro}$ > 1, the positive charge varies with pH: it follows the behavior of the net charge of lysozyme that decreases with the pH (+17 at pH 3, +11 at pH 4.7, + 8 at pH 7). In a remarkable way, we obtain the same value of |ζ| at pH 4.7 for fully and partially sulfonated chains. In summary ζ is directly linked to the amphoteric properties of the lysozyme when in complete interaction with solvent. At variance with PSS, lysozyme is excess is not localised by SANS: due to its compact shape, it does not form a sufficiently thick shell. However, the excess charge behaves exactly as if at the outside of the globule. In summary, although zetametry, which is not a localisation technique, does not distinguish between inner and outer charges, the most consistent picture is the following: the inner core is perfectly neutral, and excess species feel exactly the same pH as if immersed in the surrounding solvent, because they are localized in the vicinity of the globule surface.

Moreover, all charge values agrees well with our analysis of the composition of the primary globular complexes from SANS."

## Discussion: why a finite size in the system?

The overview of all the experiments presented in section 3 enables us to get an accurate description of the influence of the different physico-chemical parameters (pH, sulfonation rate, concentration of species, ionic strength) on the structure of PSS/lysozyme complexes in the regime of dense globules, i.e. when the PSS chains are in diluted regime after interaction with the lysozyme. This enables us to propose a precise picture of the mechanisms of the complexation process.

The first striking result is the predominance of the direct electrostatic attractions between the species in the system. The structure of dense complexes with a neutral core is recovered whatever the conditions tested ([-]/[+]$_{inner}$ ~ 1). The respective influence of the hydrophobic interactions (tested in our experiments with a sulfonation rate f = 0.5) and of the patch distribution of the charges on the proteins (tested in our experiments when the pH is 7) are thus of the second order compared to these direct electrostatic interactions since they only have a very minor influence on the inner structure of the complexes. It is particularly remarkable that the compacity $\Phi_{inner}$ stays constant with a value of ~ 0.3 for all experiments. This shows that the main mechanism of formation of complexes we described previously[28,41] is kept whatever the way the strength of electrostatic interactions are tuned. It agrees with a driving of the complexation by (i) the gain of enthalpy due to the formation of electrostatic bonds associated with (ii) the gain of entropy due to the release of the counterions, which overbalance the loss of conformational entropy of the chains. This leads to the formation of dense globular complexes with a neutral core that aggregate at large scale like charged colloids (Reaction Limited Colloidal Aggregation).

The strong linear charge density of PSS is here the key parameter in the complexation process. The distance between two charges on the chain for f=1 (~ 2 Å) is lower that the Bjerrum length (~ 7 Å), so that there is always a large reservoir of counterions condensed on the chains (Manning condensation) that can be potentially released during PSS/lysozyme complexation. For partial sulfonation f = 0.5, pearls are formed which also condense counterions.[49,50] Complexation can thus be entropically driven in all cases of this paper.

Since electrostatic interactions are prevalent in the system, they are necessarily the key point to answer the crucial question of the finite size, $R_{mean}$, of the globules. The mean number of proteins per primary complex $N_{lyso\_comp}$ (that mainly depends over $R_{mean}$ by its third power since $\Phi_{inner}$ is approximately constant) varies indeed with all the parameters that may influence such electrostatic interactions (pH, [-]/[+]$_{intro}$, species concentration, ionic strength of the buffer).

Let us now explain how $N_{lyso\_comp}$ is tuned by the electrostatic interactions. We showed in Fig. 4.d that the formation of the primary (i. e. globular) complexes is a nucleation-growth process. The growth of the primary complexes starts with a given number of germs, which then coalesce and stop growing at a finite size with $N_{lyso\_comp}$ lying from a few tens to a few hundreds. **The electrostatic interactions, which drive complexation, limit thus also the growth of primary complexes.** The electrostatic interactions should thus be considered at two different length scales: a local scale, characteristic of the size of the components- lysozyme and PSS chain, and a larger scale, characteristic of the size of the primary complexes, $R_{mean}$. On such large scales, the 2.1 fractal dimension of the aggregates of primary complexes, which is the signature of the RLCA aggregation, suggests that there are electrostatic repulsions between complexes. This is consistent with the zetametry measurements of Fig. 7 that show that the primary complexes bear a positive global charge when [-]/[+]$_{intro}$ < 1 and a global negative charge when [-]/[+]$_{intro}$ > 1. This global charge comes from the uncomplexed charged species at the surface of the primary complexes. The surface-charged primary complexes behave thus like charged colloids and interact via long-ranged electrostatic repulsions. The structure of the complexes at large scale must be strongly modulated by the screening of such electrostatic repulsions by all the ions of the solution. Classically, the typical range of the screening is the Debye length $\lambda_D$ of the solution which is defined by:

$$\frac{1}{\lambda_D} = \kappa = \sqrt{\frac{2e^2 I}{\varepsilon_r \varepsilon_0 k_B T}} \quad (5)$$

where κ is the screening constant, $e$ the elementary charge of an electron, $\varepsilon_0$ the permittivity of the vacuum, $\varepsilon_r$ the relative permittivity of the medium (78.5 in water), $k_B$ the Boltzmann constant and T the temperature.

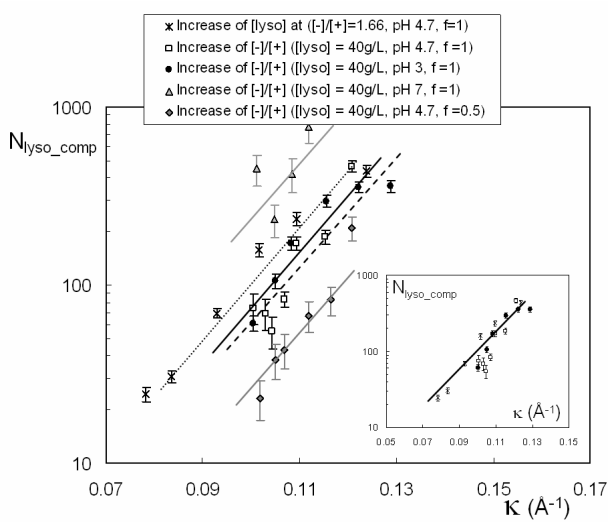

**Fig. 8** Overview of the influence of κ on mean aggregation per primary complexes $N_{lyso\_comp}$. Insert: Same plots when keeping only the samples for which electrostatic interactions are very similar (pH 3 and 4.7, f=1; see text). The lines are only guides for the eyes. Results for pH 4.7 and f = 1 are issued from a former paper.[28]

The ionic strength I in formula 5 should take into account all the co-ions and counterions present in the solvent, including the ones released from the species during complexation and the ones from the negatively charged patches of the proteins.[54] We present in Fig. 8 $N_{lyso\_comp}$ as a function of κ for all the experiments presented in the paper. Whatever the parameter tuned to change the interactions (pH, [-]/[+]$_{intro}$, concentration of the species, sulfonation rate), **$N_{lyso\_comp}$ is linear with κ in a linear-log representation**. If ones removes the experiments where hydrophobic interactions are present (pH 4.7, f = 0.5) or where electrostatic patches on the protein surface have a strong importance (pH 7, f =1), **all points lie on a master curve** (see insert of Fig 8). Let us compare these results with the ones obtained when the initial ionic strength of the buffer is 0.5 mol/L in the conditions of pH 4.7, [lyso] = 40g/L and [-]/[+]$_{intro}$= 3.33. As explained before, the sizes of the primary complexes are in this case too large to be correctly fitted in the available q-range of our experiment. But an interpolation of $N_{lyso\_comp}$ based on an $R_{mean}$ lying between 300Å of 400Å is possible. It would give an $N_{lyso\_comp}$ of a few thousands. The corresponding point (κ is 0.25 for this sample) would still lie on the master curve. The electrostatic interactions completely determine the finite size of the primary complexes since $N_{lyso\_comp}$ scales exponentially with the inverse of the Debye length κ. Moreover, when other interactions are present (like for pH 7, or f=0.5), the scaling keeps the same exponent but the curve is shifted.

This could be explained as follows. Like in the classical DLVO theory of charged colloids [55], we write the repulsive electrostatic potential $V_{el}(r)$ between two primary complexes each of charge Z, as a Yukawa potential:

$$\frac{V_{el}(r)}{kT} = Z_{comp}^2 L_B \frac{e^{-\kappa r}}{r} \quad (6)$$

where $Z_{comp}$ is the global charge of a primary complex and $L_B$ the Bjerrum length.

If other forces than pure electrostatic can be described by an additional potential V'(r) (positive or negative). So that the long-repulsions between primary complexes $V_{rep}(r)$ write:

$$\frac{V_{rep}(r)}{kT} = \frac{V_{el}(r)}{kT} + \frac{V'(r)}{kT} \quad (7)$$

An exact calculation of $V_{el}(r)$ would require to determinate precisely the exact value of $Z_{comp}$, which is not possible by zetametry. But even without knowing this exact value, the effect of $V_{rep}(r)$ on $N_{lyso\_comp}$ can be determined. Since the core of the primary complexes is neutral, $Z_{comp}$ is a surface charge:

$$Z_{comp} \propto R_{mean}^2 \propto N_{lyso\_comp}^{2/3} \quad (8)$$

$V_{el}(r)$ scales thus like $N_{lyso\_comp}^{4/3}$ and increases strongly during the growth of the primary complexes for a given ionic strength. When germs of few proteins are formed at the beginning of the complexation, the maximum of $V_{rep}(r)$ –i. e. the activation barrier- is negligible compared to $k_BT$. The complexation is here driven by electrostatic attractions associated with the charges of the species on a very short range (there should indeed remain some charged species of opposite sign compared to the global charge of the complexes at its surface). When the maximum of $V_{rep}(r)$ reaches a value $V_{rep}^*$ of a few $k_BT$, the long-range repulsions between complexes become dominating in the system. The aggregation growth crosses over then from a strong collapse between species of opposite charge to the RLCA aggregation process of charged colloids, as observed in q space and real space.[41] $Z_{comp}$ thus no longer increases.

$V_{rep}^*$ has a similar form for all samples. When there are only electrostatic interactions in the system, it can be directly derived from equation (6). It is maximum at contact between aggregates of radius $R_{mean}$ and we get:

$$\frac{V_{rep}^*}{kT} \propto N_{lyso\_comp}^{4/3} L_B \frac{e^{-2\kappa R_{mean}}}{2R_{mean}} \quad (9)$$

Since $N_{lyso\_comp} \sim R_{mean}^3 \phi_{lyso\_inner}$,

$$\frac{V_{rep}^*}{kT} \propto R_{mean}^4 \phi_{lyso\_inner}^{4/3} L_B \frac{e^{-2\kappa R_{mean}}}{2R_{mean}} \quad (10)$$

Let us take a sensible value for $V_{rep}^*$, say 3 kT, Assuming that V* stays around this value implies that the second member of Eq. (10) is constant, i.e. $R_{mean}$ is a given function of κ only and so does $N_{lyso\_comp}$.

When other interactions are present in the system, they only shift the curve of figure 8 if they do not depend on the size of the primary complexes. If these other interactions are attractive, they shift $N_{lyso\_comp}$ toward lower values for a given κ. This is the case when hydrophobic interactions can be present like for f = 0.5 (see Fig 8). The presence of negatively charged patches induces the opposite effect (see the results for pH 7).

Once the primary complexes are formed, their organization at larger scales is controlled by their surface charge $Z_{comp}$. Aggregation occurs when the thermal motion is able to overcome the threshold $V_{rep}^*$. This is determined by the range of the attractive part of the interaction potential. When this potential is long ranged, coacervation (i.e. demixion between rich and poor phases of dense complexes[43], which is a 'gas-liquid' transition in its principle, and a liquid-liquid phase separation in practice) could occur in the system as predicted on colloids.[56] This is typical of weakly charged objects and can thus be observed with weakly charged polyelectrolytes such as polysaccharides.[3] When this potential is short range, i.e for strongly charged objects, aggregation between objects simply occurs. Since the PSS chains are strongly charged, $Z_{comp}$ is high and the mechanism of aggregation in our system is RLCA aggregation rather than coacervation. Finally, all the experiments presented all along the manuscript enable us to picture the precise structures of the globules as a function of the different physical chemical parameters of the system. They are recalled in Fig 9.

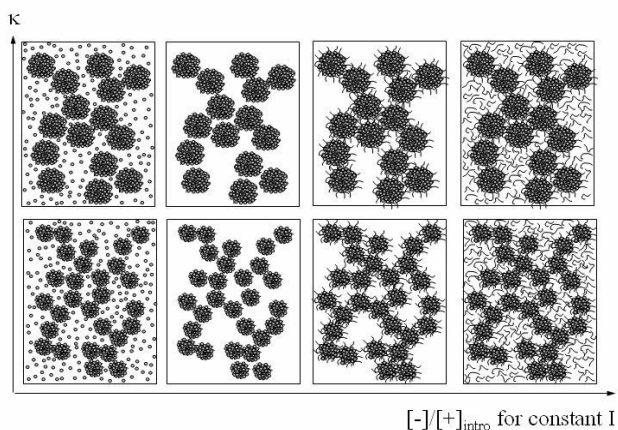

**Fig. 9** Pictures of the different structures of the complexes. Please note that [-]/[+]intro depends both on the pH and on the sulfonation rate.

## Conclusions

We have presented here a complete study of the complexation of a positively charged protein (lysozyme) with a negatively charged polyelectrolyte chain (PSS) by testing the influence of all the parameters that may act on the structure of the complexes (charge densities and concentration of the species, partial hydrophobicity of the polyion chain, ionic strength), in a regime where the PSS chains stay in dilute regime after interaction with the lysozyme. Whatever the conditions, the same kinds of structures are formed on the scales lying from 10Å to 1000 Å: the complexation leads to the formation of dense globules of a few hundreds Å with a neutral core ([-]/[+]$_{inner}$ ~ 1) and a compacity of ~ 0.3. The species in excess from an electrostatic point of view are not localized in the core. They remain free in solution, except a fraction of them when [-]/[+]$_{intro}$ is slightly superior to 1. In the latter case, some PSS chains in excess are located first in a PSS shell. Due to the presence of charged species at their surface, the primary complexes have a positive charge when [-]/[+]$_{intro}$ > 1 and a negative charge when [-]/[+]$_{intro}$ < 1.

Two particularly striking results show a very general pregnancy of electrostatics:

The inner charge ratio obtained from SANS shows a master behavior as a function of introduced charge ratio (Fig 3). This is confirmed by zetametry, which supports the picture of excess charge localized at the surface.

The size of the globules shows a common behavior as a function of the screening length $\kappa^{-1}$ calculated from the actual ionic strength (Fig 8).

The primary complexes thus resemble charged colloids. Since we know their size, have a relative estimate of their charge, and electrostatics screening seems to be the control parameter, we propose to relate these three parameters to give an origin of the finite size.

We propose that this size is determined by the crossover between two types of aggregation. During an initial stage of nucleation and growth of globules driven only by electrostatic attractions associated with counterion release, the global charge increases, increasing long-ranged electrostatic repulsions between the primary globules. Repulsions finally become dominent, which stops the growth of the complexes at a finite size. The complexes start then to aggregate like charged colloids with a RLCA process, which leads to the formation of fractal clusters of primary complexes of dimension 2.1. **The finite size is completely driven by the electrostatic interactions because the mean number of proteins per primary complex N$_{lyso\_comp}$ grows exponentially on a master curve with the inverse of the Debye length $\kappa$** whatever the way the physicochemical parameters are modified in the system. The variation of N$_{lyso\_comp}$ with $\kappa^{-1}$ is similar and only shifted by differences in interactions: at partial sulfonation, related to the introduction of hydrophobic forces or at high pH 7, probably due to the presence of patches of negative charges on the protein.

The structural properties of the primary complexes (size, surface charge and/or hydrophobicity) can be thus completely determined by the synthesis conditions in a reproducible way. This enables a direct tuning of the properties of the solutions such as colloidal stability or adhesion, which is very promising for potential applications. Moreover, it is possible to tune the density of enzymes (here lysozyme) remaining at the surface, and hence keeping their enzymatic activity.

Nevertheless, there is a still pending crucial question. It concerns the potential reorganizations that could occur in the system either kinetically or under the action of an external stimulus. This is a key point for potential applications. In this paper, we have clearly established that there are two typical scales for the interactions in the system (within the globules and between the globules) with different strengths and ranges. This must have a strong impact on the capacity of reorganizations of the complexes. We will thus investigate the possibility of reorganizations of the complexes at the different relevant scales of the system in a forthcoming paper.

## Acknowledgments

The authors are greatly indebted to Monique Axelos and Dominique Guibert (BIA, INRA Nantes, France) for giving us access to their zetameter and help during measurements.